\documentclass[aps,prl,reprint,amsmath,amssymb,floatfix]{revtex4-2}

\usepackage[utf8]{inputenc}
\usepackage{graphicx}
\usepackage{dcolumn}
\usepackage{bm}
\usepackage{lipsum}
\usepackage[hidelinks]{hyperref}

\renewcommand{\vec}[1]{\mathrm{\mathbf{#1}}}

\begin{document}

\title{Bayesian MINFLUX localization microscopy}

\author{Steffen Schultze}
\affiliation{Max Planck Institute for Multidisciplinary Sciences}
\author{Helmut Grubmüller}
\email{hgrubmu@mpinat.mpg.de}
\affiliation{Max Planck Institute for Multidisciplinary Sciences}

\date{\today}

\begin{abstract}
MINFLUX microscopy allows for localization of fluorophores with nanometer precision using targeted scanning with an illumination profile with a minimum. However, current scanning patterns and the overall procedure are based on heuristics, and may therefore be suboptimal. Here we present a rigorous Bayesian that offers maximal resolutions from either minimal detected photons or minimal exposures. We estimate using simulated localization runs that this approach should reduce the number of photons required for $1\,\mathrm{nm}$ resolution by a factor of about four.  
\end{abstract}

\maketitle
Recently, MINFLUX experiments \cite{balzarotti_nanometer_2017,eilers_minflux_2018,gwosch_minflux_2020,schmidt_minflux_2021,sahl_direct_2024,scheiderer_minflux_2025} have pushed the resolution limit of optical fluorescence microscopy into the nanometer range. 
They also allow for particle tracking using a minimal number of detected photons \cite{balzarotti_nanometer_2017, scheiderer_minsted_2024}.
This high photon efficiency is achieved by localizing single fluorophores using a structured illumination profile with a central intensity minimum, such as a donut-shaped intensity profile.
By placing this minimum at a small number of positions near the near the fluorophore to be localized and recording photon counts for each, the fluorophore location can be estimated with nanometer accuracy. 
Ideally, if the position were `guessed' accurately and assuming negligible background noise, localization would be possible without even a single detected photon.

In practice, localization is performed iteratively \cite{balzarotti_nanometer_2017}. 
Starting from a diffraction limited pre-localization at about $150\,\mathrm{nm}$ precision, the donut minimum is positioned at the vertices of a triangular or hexagonal scanning pattern centered on the current position estimate. The pattern diameter $L$ is chosen according to the initial localization precision. 
Photon counts at these positions are used to update the emitter estimate, after which the scanning pattern is re-centered and its diameter reduced. 
This procedure progressively directs the excitation minima toward the true emitter position. 
Because the effective localization precision scales linearly with the pattern diameter $L$, the iterative refinement yields exponential accuracy improvement with respect to the number of detected photons, in contrast to the slower $1/\sqrt{N}$ scaling obtained without such repositioning.

However, this overall procedure as well as the triangular or hexagonal scanning patterns are, essentially, heuristics.
Therefore, and because not all available information is used, they are likely suboptimal, both in terms of the required number of detected photons as well as the tolerance to background noise due to a non-zero intensity minimum.

Here we develop a Bayesian scanning strategy that is optimal, either regarding the required number of detected photons or the number of exposures, such that it achieves either maximal resolutions from a given photon budget, or maximal time resolution.
In each step we ask at which position of the donut minimum the expected information gain is maximal. 
As we will demonstrate, this approach greatly improves the photon efficiency of each localization, achieving, for example, a typical target resolution of $2\,\mathrm{nm}$ using four times fewer photons than required in current implementations.

\begin{figure*}
    \includegraphics[width=\linewidth,trim={0.4cm 0 0 0}]{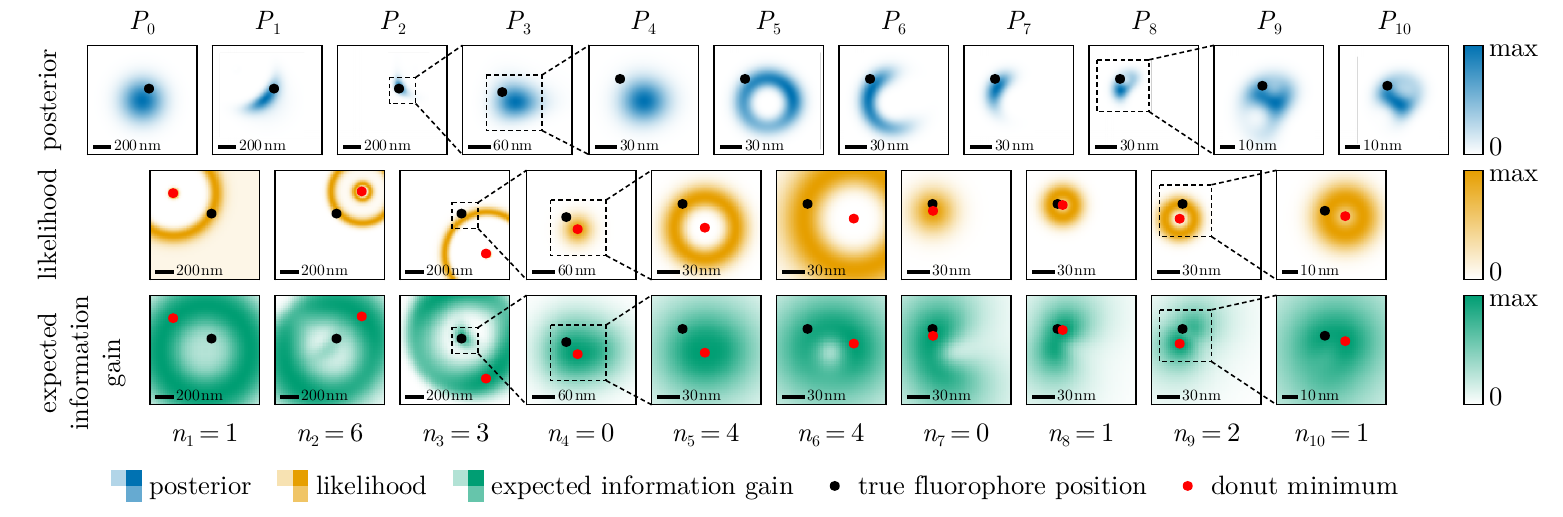}
    \caption{Selected Bayesian MINFLUX localization. Shown are the posterior densities $P_k$ (blue), the likelihoods $P(n_k \mid \vec{x}, \vec r_k, \eta_k)$ (orange), the expected information gain $\mathrm{EIG}(\vec r, P_{k-1})$ (green), and the observed photon counts $n_k$.
    Localization was performed from an isotropic Gaussian prior $P_0$ with standard deviation $\sigma = 150\,\mathrm{nm}$ with an expected photon count per step of $\mu = 2$ and minimum positions (red dots) chosen to maximize the expected information gain.}
    \label{fig: example}
\end{figure*}

\emph{Bayesian framework.}
We treat the localization as a sequential Bayesian inference problem. To that end, let $P_0(\vec{x})$ denote the prior distribution for the emitter position $\vec x$, obtained from an initial diffraction-limited pre-localization.
As illustrated in Fig.~\ref{fig: example}, we separately consider each exposure $k = 1,\dots,N$ with excitation profiles $I_k(\vec{x}) = I(\vec{x}; \vec{r}_k, \eta_k)$ determined by independently chosen minimum positions $\vec r_k$ (red dots in Fig.~\ref{fig: example}) and intensity factors $\eta_k$. Here, 
\begin{equation}
    I(\vec x; \vec r, \eta) = \eta \left[e(1-b) \frac{\lVert \vec x - \vec r \rVert^2}{\sigma^2} \exp\left(-\frac{\lVert \vec x - \vec r \rVert^2}{\sigma^2}\right) + b \right]
\end{equation}
is the donut-shaped illumination profile, and $b$ is the background level, $\sigma \approx 200\,\mathrm{nm}$ is radius of the donut maximum and $e$ is Euler's number.

In each step, both $\vec r_k$ and $\eta_k$ are chosen based on the current Bayesian posterior $P_k(\vec x)$ (blue density in Fig.~\ref{fig: example}), as explained below.
After observing the corresponding photon count $n_k$, the posterior is updated according to Bayes’ theorem,
\begin{equation}\label{eq: bayes}
    P_k(\vec{x}) \propto P(n_k \mid \vec{x}, \vec r_k, \eta_k)\, P_{k-1}(\vec{x}),
\end{equation}
such that, after each step, $P_k(\vec x)$ encodes all information from the steps up to this point.
Modeling each photon count $n_k$ as an independent Poisson distribution $n_k \sim \mathrm{Pois}(I_k(\vec x_0))$, where $\vec x_0$ is the ground-truth dye position, the likelihood (orange density in Fig.~\ref{fig: example}) for a given exposure reads
\begin{equation}
    P(n_k \mid \vec{x}, \vec r_k, \eta_k) = \frac{e^{-I_k(\vec{x})} \, I_k(\vec{x})^{n_k}}{n_k!}.
\end{equation}

\emph{Optimal donut placement.}
Crucially, the choice of the next minimum position $\vec{r}_k$ is not fixed but adapted in each step based on the current posterior $P_{k-1}(\vec{x})$.
Indeed, having access to this posterior allows to estimate the expected information gain for each possible donut placement $\vec r$. 
To avoid a bias towards positions with higher expected photon counts (which will yield more information on average, but less information per photon), for each possible $\vec r_k$ the intensity factor $\eta_k$ is chosen such that the expected photon count given $P_{k-1}$ equaled a chosen parameter $\mu$, by setting $\eta_k = \eta(\vec r_k)$ with 
\begin{equation}
    \eta(\vec r) = \left(\mathbb E_{\vec x \sim P_{k-1}(\vec x)} I(\vec x; \vec r, 1)\right)^{-1}\mu.
\end{equation}

\begin{figure}
    \includegraphics[width=\linewidth]{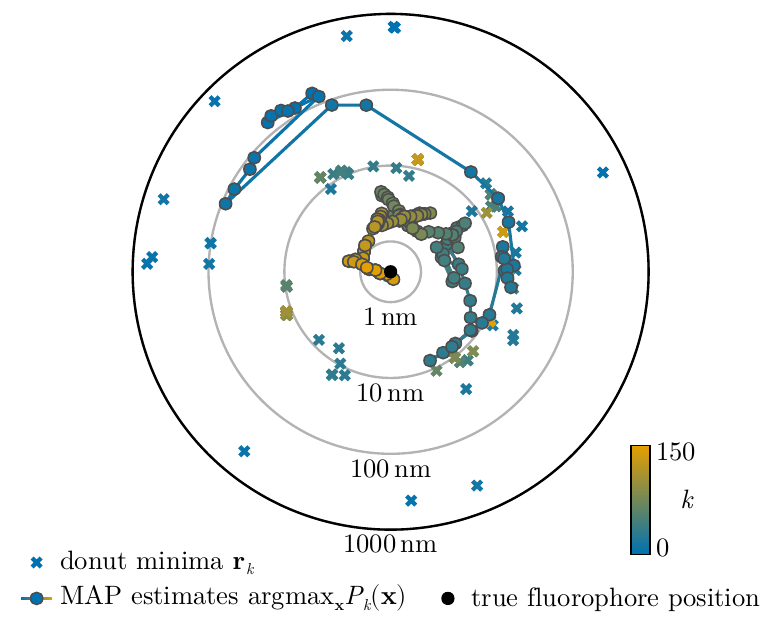}
    \caption{Current maximum a posteriori (MAP) estimates $\operatorname{argmax}_{\vec x} P_k(\vec x)$ and donut minimum positions $\vec r_k$ for a selected Bayesian MINFLUX localization run performed with $\mu = 0.5$ for $150$ exposures $k$. Radial scale is linear below $1\,\mathrm{nm}$ and logarithmic above.}
    \label{fig: exampletrace}
\end{figure}

Assuming an observed photon count $n$, the information gain is defined as the reduction in entropy $S(P) = \int P(\vec x) \log P(\vec x)\,\mathrm d\vec x$ from $P_{k-1}(\vec x)$ to $P_k(\vec x \mid n) \propto P(n \mid \vec{x}, \vec r_k, \eta_k)\, P_{k-1}(\vec{x})$.
Taking the expectation over all possible photon counts yields the expected information gain
\begin{align}\label{eq: eig}
    \mathrm{EIG}&(\vec r, P_{k-1}) \\
    &= \mathbb E_{n \sim P(n \mid\vec r)} [S[P_{k-1}(\vec x)] - S[P_k(\vec x \,|\, n)]] \nonumber\\
    &= S[P_{k-1}(\vec x)] - \sum_{n=0}^\infty P(n\,|\, \vec r) \,S[P(\vec x\,|\, n, \vec r, \eta(\vec r))]],  \nonumber
\end{align}
as shown in the bottom row of Fig.~\ref{fig: example}. Here,
\begin{equation}
    P(n \,|\, \vec r) = \int \mathrm{Pois}(n \,|\, I(\vec x; \vec r, \eta(\vec r))) P_{k-1}(\vec x)\, \mathrm d \vec x.
\end{equation}
denotes the expected probability of each photon count $n$ given $P_{k-1}$. 
The optimal $\vec r_k$ is now chosen as the one that maximizes the EIG, 
\begin{equation}\label{eq: argmaxeig}
    \vec r_k = \operatorname{argmax}_{\vec r} \mathrm{EIG}(\vec r, P_{k-1}).
\end{equation}
Figure~\ref{fig: exampletrace} shows a selected simulated localization run using this optimal choice of $\vec r_k$. 

\begin{figure}
    \includegraphics[width=\linewidth]{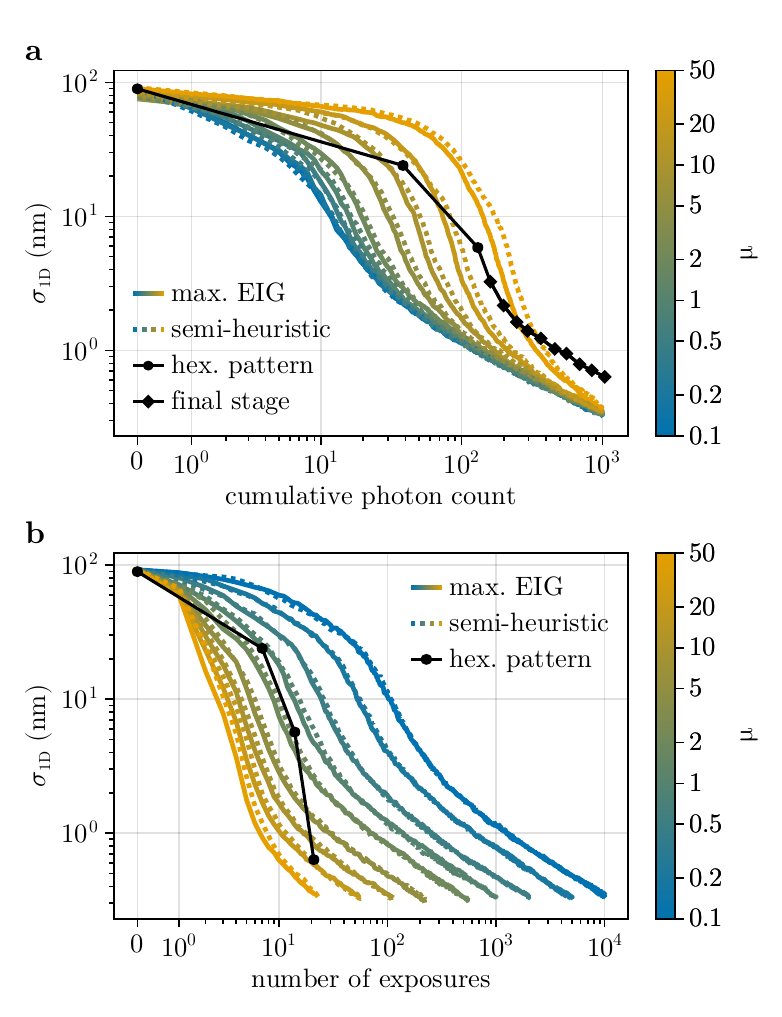}
    \caption{Median localization accuracy $\sigma_\mathrm{1D}$ for various values of $\mu$ as function of \textbf{a} the cumulative detected photon count and \textbf{b} the number of exposures for the Bayesian approach using maximized EIG (solid colored lines) and semi-heuristic placement (dotted lines) as well as the conventional approach using a hexagonal pattern (black line) with varying photon counts for the final stage (diamond markers).}
    \label{fig: photonefficiency}
\end{figure}

\emph{Photon efficiency.}
To assess the efficiency of this optimized Bayesian approach in comparison to the state-of-the-art heuristic approach, we simulated $1000$ independent localization runs for various values for the average photon count per step $\mu$, as shown in Fig.~\ref{fig: photonefficiency} (solid~lines).
Each run was started from a Gaussian prior with standard deviation $150\,\mathrm{nm}$, from which the ground-truth emitter position was chosen at random. 
For each step, the maximum expected information gain donut minimum position was chosen. 
The posterior was represented by a discretization on a two-dimensional grid of points and the required integrals where computed numerically, as explained in the Appendix. 
A realistic background level of $b = 0.01$ was chosen, corresponding to signal to background ratio of $12$ at a $L = 100\,\mathrm{nm}$ \cite{balzarotti_nanometer_2017}.
Localization accuracy was measured by the median of the one-dimensional errors \cite{schmidt_minflux_2021} of the maximum a posteriori (MAP) estimates given by $\sigma_{1\mathrm D} = \lVert \vec x_\mathrm{MAP} - \vec x_0\rVert/\sqrt{2}$, where $x_\mathrm{MAP} = \operatorname{argmax}_\vec x P_k(\vec x)$. 
As can be seen in Fig.~\ref{fig: photonefficiency}a and b, choosing the expected photon count per step presents a tradeoff between using minimal photon counts or minimal numbers of steps. 

\begin{figure}
    \includegraphics[width=\linewidth]{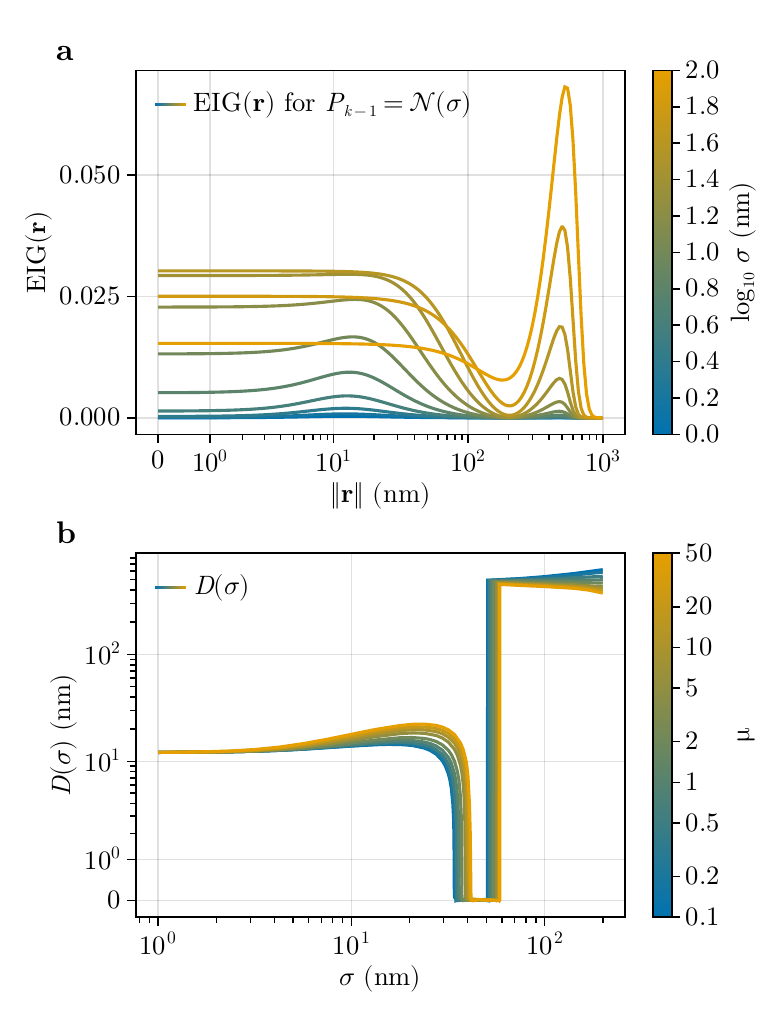}
    \caption{Optimization of $\vec r_k$ for $P_{k-1} = \mathcal{N(\sigma)}$ at $\mu = 0.1$ photons per step. \textbf{a} Expected information gain as a function of $\lVert \vec r\rVert$ for various values of $\sigma$. \textbf{b} Distance $D(\sigma) = \lVert \operatorname{argmax}_{\vec r} \mathrm{EIG}(\vec r, \mathcal N(\sigma))\rVert$ at which the expected information gain is maximized as a function of $\sigma$ for various values of $\mu$.}
    \label{fig: eigsforgaussian}
\end{figure}

Strikingly, for $\mu = 0.1$ already about $120$ photons suffice to reach $1\,\mathrm{nm}$ average accuracy. 
In contrast, using the standard iterative maximum likelihood approach (black lines in Fig.~\ref{fig: photonefficiency}) about $500$ photons are required to reach the same accuracy.
For this comparison, we simulated $1000$ localization runs using a hexagonal pattern with three stages of $L_1 = 200\,\mathrm{nm}$, $L_2 = 150\,\mathrm{nm}$ and $L_3 = 40\,\mathrm{nm}$ at reported photon counts of $N_1 = 40$ and $N_2 = 90$ for the first two stages \cite{schmidt_minflux_2021} and a range photon counts $N_3$ for the final third stage (diamond markers in Fig~\ref{fig: photonefficiency}).

The conventional approach achieved slightly better resolutions in our simulations than reported from experiments \cite{schmidt_minflux_2021}, likely due to higher experimental noise-levels or other additional sources of uncertainty. 
The relative performance of both approaches should not be affected by this difference. 
Notably, the Bayesian approach is more photon-efficient than the standard method also for some of the larger tested values of $\mu$. 
Indeed, for about $\mu = 2$, for which the Bayesian approach requires a similar amount of exposures than the standard method, $1\,\mathrm{nm}$ accuracy is still reached using nearly the same amount of detected photons as for $\mu = 0.1$.

Somewhat unexpectedly, at the beginning of the localization process the most efficient donut placement is not obtained by aligning the intensity minimum with the region of highest posterior probability. 
Instead, maximal information is gained when the outer slope of the excitation profile overlaps with the support of the posterior distribution. 
Only once the posterior has become sufficiently narrow does positioning the minimum itself close to the emitter become optimal. 
To understand this behavior, we computed the expected information gain as a function of the radial distance $\lVert \vec r\rVert$ of the donut minimum, assuming $P_{k-1} = \mathcal{N}(\sigma)$ is an isotropic Gaussian centered at the origin with standard deviation $\sigma$ (Fig.~\ref{fig: eigsforgaussian}a). 
For large $\sigma$, the EIG is maximized at large radial offsets, consistent with the outer donut slope being most informative. This is further illustrated in Fig.~\ref{fig: eigsforgaussian}b, which shows the optimal placement distance
\begin{equation}
    D(\sigma) = \lVert \operatorname{argmax}_{\vec r} \mathrm{EIG}(\vec r, \mathcal N(\sigma))\rVert
\end{equation}
as a function of $\sigma$. 
Notably, due to the background $b$, for narrow priors the optimal distance does not become zero but instead approaches a finite offset, consistent with earlier results on the optimal scanning pattern size $L$ in conventional MINFLUX \cite{eilers_minflux_2018}.

Regarding the implementation of our approach into existing microscopes, maximizing the EIG requires considerable computational effort due to the required numerical evaluation of integrals. 
As MINFLUX-localizations need to be performed within a few milliseconds, this computational cost might become a bottleneck. 
Conjecturing that the radial distance of the minimum position to the current maximum likelihood position is the most important factor, we therefore suggest a semi-heuristic placement strategy as
\begin{equation}
    \vec r_k = \operatorname{argmax}_{\vec x} P_{k-1}(\vec x) + \vec v_k D(\operatorname{std}(P_{k-1})),
\end{equation}
where $\vec v_k$ is an isotropic random unit vector and $D(\operatorname{std}(P_{k-1}))$ is the optimal distance from Fig.~\ref{fig: eigsforgaussian}b. 
This placement strategy is similar to one used in combination with stimulated emission depletion microscopy (MINSTED) \cite{weber_minsted_2021,scheiderer_minsted_2024}, but differs in that it is here used in combination with updating the Bayesian posterior instead of a point estimate.
As can be seen in Fig.~\ref{fig: photonefficiency} (dotted lines), our semi-heuristic placement strategy performs nearly as well as the rigorous maximum EIG strategy.

\begin{figure}
    \includegraphics[width=0.49\textwidth]{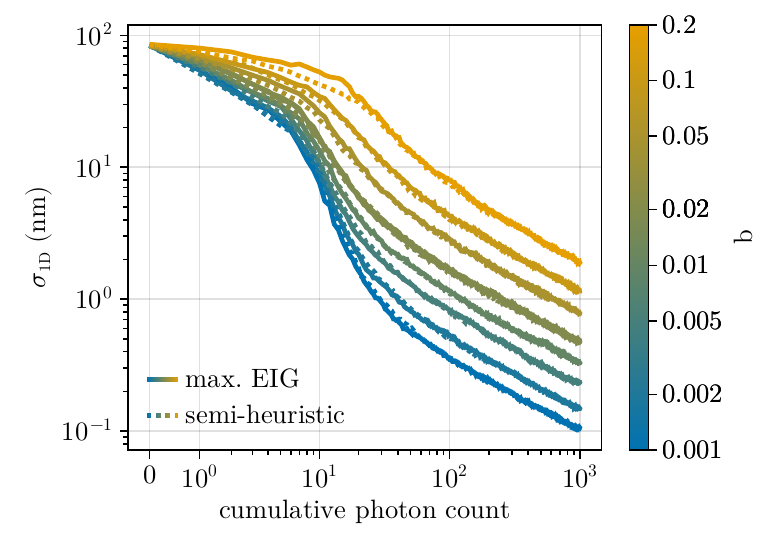}
    \caption{Median localization accuracy $\sigma_\mathrm{1D}$ for various background levels $b$ as function of the cumulative detected photon count for both optimized EIG (solid lines) and radial heuristic placement (dashed lines).}
    \label{fig: background}
\end{figure}

\emph{Background tolerance.}
The background level $b$ is the most important limitation on achievable resolutions from MINFLUX. 
Indeed, as can be seen in Fig.~\ref{fig: photonefficiency}, at about $1\,\mathrm{nm}$ accuracy the further increase in accuracy changes from exponential to inverse square root behavior because the background becomes dominant at this scale.
As the background level may vary greatly depending on experimental setup and fluorophore type, we also investigated how our approach performs at elevated background levels. 
To that end, we performed another $1000$ optimization runs for several background levels ranging from $b = 0.001$ to $b = 0.2$, corresponding to signal to background ratios of $120$ and $0.5$ at $L = 100\,\mathrm{nm}$ (Fig.~\ref{fig: background}).
As expected, at higher noise levels more photons are required for the same accuracy. 
Notably, at $5$ times increased noise level ($b = 0.05$) our approach is as efficient as the conventional one at typical noise levels ($b = 0.01$).

\emph{Conclusion.} 
Here we have developed and assessed a Bayesian approach to MINFLUX localization microscopy, which, by construction, yields an optimal scanning strategy for positioning the donut intensity minimum.
Our simulated localization runs suggest that this approach should provide similar accuracy using four times fewer photons than current state-of-the-art approaches. 
Conversely, with unlimited numbers of photons, the time resolution, that is, the required number of exposures, should improve by about a factor of three.
We attribute this increase in efficiency not only to the optimized donut minimum positions, but also to the fact that our approach uses and maintains all available information.
While our approach provides optimal single-step information gain, it may be possible to improve even further by considering the expected information gain for multiple steps at once. 
Preliminary tests using the two-step EIG suggest a further $10\%$ reduction in required photon count, albeit at markedly higher numerical effort. 

Whereas we have here for simplicity assumed exact knowledge of the fluorophore brightness as well as the shape of the illumination profile, these are often not exactly known in practice. 
Importantly, the Bayesian framework can similarly be applied to simultaneously determine both the fluorophore position and these parameters. 
For instance, for the fluorophore brightness, an overall intensity factor $\eta$ would be included and the updates performed on a joint posterior
\begin{equation*}
    P_k(\vec{x}, \eta) \propto P(n_k \mid \vec{x}, \eta, I_k)\, P_{k-1}(\vec{x}, \eta).
\end{equation*}

Our approach is computationally more demanding than established methods. 
However, while its implementation on current field-programmable gate arrays used in MINFLUX setups may be challenging, it should be feasible using upgraded computational hardware. 

We have here focused on the most commonly used donut-shaped excitation intensity, MINFLUX can also be applied using alternative illumination profiles \cite{wirth_minflux_2023, deguchi_simple_2024} with, for example, line-shaped minima \cite{wirth_minflux_2023}.
Further analysis will show if an analogous Bayesian approach would offer similar improvement also for these alternatives.
Notably, in the case of a line minimum the localization is independent for each dimension, such that the computational complexity would be much lower. 

\vspace{2ex}
\emph{Acknowledgments}. 
We thank Stefan Hell, Marcel Leutenegger, Jan Otto Wirth, and Johann Engelhardt for helpful discussions. This research was inspired by listening to a presentation by Stefan Hell.
This work was supported by the Deutsche Forschungsgemeinschaft (DFG, German Research Foundation) - CRC 1456/1 - 432680300.

\vspace{2ex}
\emph{Data availability}.
The data that support the findings of this article are available from the authors upon reasonable request. 

\bibliography{main.bib}

\begin{thebibliography}{12}%
\makeatletter
\providecommand \@ifxundefined [1]{%
 \@ifx{#1\undefined}
}%
\providecommand \@ifnum [1]{%
 \ifnum #1\expandafter \@firstoftwo
 \else \expandafter \@secondoftwo
 \fi
}%
\providecommand \@ifx [1]{%
 \ifx #1\expandafter \@firstoftwo
 \else \expandafter \@secondoftwo
 \fi
}%
\providecommand \natexlab [1]{#1}%
\providecommand \enquote  [1]{``#1''}%
\providecommand \bibnamefont  [1]{#1}%
\providecommand \bibfnamefont [1]{#1}%
\providecommand \citenamefont [1]{#1}%
\providecommand \href@noop [0]{\@secondoftwo}%
\providecommand \href [0]{\begingroup \@sanitize@url \@href}%
\providecommand \@href[1]{\@@startlink{#1}\@@href}%
\providecommand \@@href[1]{\endgroup#1\@@endlink}%
\providecommand \@sanitize@url [0]{\catcode `\\12\catcode `\$12\catcode
  `\&12\catcode `\#12\catcode `\^12\catcode `\_12\catcode `\%12\relax}%
\providecommand \@@startlink[1]{}%
\providecommand \@@endlink[0]{}%
\providecommand \url  [0]{\begingroup\@sanitize@url \@url }%
\providecommand \@url [1]{\endgroup\@href {#1}{\urlprefix }}%
\providecommand \urlprefix  [0]{URL }%
\providecommand \Eprint [0]{\href }%
\providecommand \doibase [0]{https://doi.org/}%
\providecommand \selectlanguage [0]{\@gobble}%
\providecommand \bibinfo  [0]{\@secondoftwo}%
\providecommand \bibfield  [0]{\@secondoftwo}%
\providecommand \translation [1]{[#1]}%
\providecommand \BibitemOpen [0]{}%
\providecommand \bibitemStop [0]{}%
\providecommand \bibitemNoStop [0]{.\EOS\space}%
\providecommand \EOS [0]{\spacefactor3000\relax}%
\providecommand \BibitemShut  [1]{\csname bibitem#1\endcsname}%
\let\auto@bib@innerbib\@empty
%</preamble>
\bibitem [{\citenamefont {Balzarotti}\ \emph {et~al.}(2017)\citenamefont
  {Balzarotti}, \citenamefont {Eilers}, \citenamefont {Gwosch}, \citenamefont
  {Gynn{\aa}}, \citenamefont {Westphal}, \citenamefont {Stefani}, \citenamefont
  {Elf},\ and\ \citenamefont {Hell}}]{balzarotti_nanometer_2017}%
  \BibitemOpen
  \bibfield  {author} {\bibinfo {author} {\bibfnamefont {F.}~\bibnamefont
  {Balzarotti}}, \bibinfo {author} {\bibfnamefont {Y.}~\bibnamefont {Eilers}},
  \bibinfo {author} {\bibfnamefont {K.~C.}\ \bibnamefont {Gwosch}}, \bibinfo
  {author} {\bibfnamefont {A.~H.}\ \bibnamefont {Gynn{\aa}}}, \bibinfo {author}
  {\bibfnamefont {V.}~\bibnamefont {Westphal}}, \bibinfo {author}
  {\bibfnamefont {F.~D.}\ \bibnamefont {Stefani}}, \bibinfo {author}
  {\bibfnamefont {J.}~\bibnamefont {Elf}},\ and\ \bibinfo {author}
  {\bibfnamefont {S.~W.}\ \bibnamefont {Hell}},\ }\bibfield  {title} {\bibinfo
  {title} {Nanometer resolution imaging and tracking of fluorescent molecules
  with minimal photon fluxes},\ }\href
  {https://doi.org/10.1126/science.aak9913} {\bibfield  {journal} {\bibinfo
  {journal} {Science}\ }\textbf {\bibinfo {volume} {355}},\ \bibinfo {pages}
  {606} (\bibinfo {year} {2017})}\BibitemShut {NoStop}%
\bibitem [{\citenamefont {Eilers}\ \emph {et~al.}(2018)\citenamefont {Eilers},
  \citenamefont {Ta}, \citenamefont {Gwosch}, \citenamefont {Balzarotti},\ and\
  \citenamefont {Hell}}]{eilers_minflux_2018}%
  \BibitemOpen
  \bibfield  {author} {\bibinfo {author} {\bibfnamefont {Y.}~\bibnamefont
  {Eilers}}, \bibinfo {author} {\bibfnamefont {H.}~\bibnamefont {Ta}}, \bibinfo
  {author} {\bibfnamefont {K.~C.}\ \bibnamefont {Gwosch}}, \bibinfo {author}
  {\bibfnamefont {F.}~\bibnamefont {Balzarotti}},\ and\ \bibinfo {author}
  {\bibfnamefont {S.~W.}\ \bibnamefont {Hell}},\ }\bibfield  {title} {\bibinfo
  {title} {{{MINFLUX}} monitors rapid molecular jumps with superior
  spatiotemporal resolution},\ }\href {https://doi.org/10.1073/pnas.1801672115}
  {\bibfield  {journal} {\bibinfo  {journal} {Proceedings of the National
  Academy of Sciences}\ }\textbf {\bibinfo {volume} {115}},\ \bibinfo {pages}
  {6117} (\bibinfo {year} {2018})}\BibitemShut {NoStop}%
\bibitem [{\citenamefont {Gwosch}\ \emph {et~al.}(2020)\citenamefont {Gwosch},
  \citenamefont {Pape}, \citenamefont {Balzarotti}, \citenamefont {Hoess},
  \citenamefont {Ellenberg}, \citenamefont {Ries},\ and\ \citenamefont
  {Hell}}]{gwosch_minflux_2020}%
  \BibitemOpen
  \bibfield  {author} {\bibinfo {author} {\bibfnamefont {K.~C.}\ \bibnamefont
  {Gwosch}}, \bibinfo {author} {\bibfnamefont {J.~K.}\ \bibnamefont {Pape}},
  \bibinfo {author} {\bibfnamefont {F.}~\bibnamefont {Balzarotti}}, \bibinfo
  {author} {\bibfnamefont {P.}~\bibnamefont {Hoess}}, \bibinfo {author}
  {\bibfnamefont {J.}~\bibnamefont {Ellenberg}}, \bibinfo {author}
  {\bibfnamefont {J.}~\bibnamefont {Ries}},\ and\ \bibinfo {author}
  {\bibfnamefont {S.~W.}\ \bibnamefont {Hell}},\ }\bibfield  {title} {\bibinfo
  {title} {{{MINFLUX}} nanoscopy delivers {{3D}} multicolor nanometer
  resolution in cells},\ }\href {https://doi.org/10.1038/s41592-019-0688-0}
  {\bibfield  {journal} {\bibinfo  {journal} {Nature Methods}\ }\textbf
  {\bibinfo {volume} {17}},\ \bibinfo {pages} {217} (\bibinfo {year}
  {2020})}\BibitemShut {NoStop}%
\bibitem [{\citenamefont {Schmidt}\ \emph {et~al.}(2021)\citenamefont
  {Schmidt}, \citenamefont {Weihs}, \citenamefont {Wurm}, \citenamefont
  {Jansen}, \citenamefont {Rehman}, \citenamefont {Sahl},\ and\ \citenamefont
  {Hell}}]{schmidt_minflux_2021}%
  \BibitemOpen
  \bibfield  {author} {\bibinfo {author} {\bibfnamefont {R.}~\bibnamefont
  {Schmidt}}, \bibinfo {author} {\bibfnamefont {T.}~\bibnamefont {Weihs}},
  \bibinfo {author} {\bibfnamefont {C.~A.}\ \bibnamefont {Wurm}}, \bibinfo
  {author} {\bibfnamefont {I.}~\bibnamefont {Jansen}}, \bibinfo {author}
  {\bibfnamefont {J.}~\bibnamefont {Rehman}}, \bibinfo {author} {\bibfnamefont
  {S.~J.}\ \bibnamefont {Sahl}},\ and\ \bibinfo {author} {\bibfnamefont
  {S.~W.}\ \bibnamefont {Hell}},\ }\bibfield  {title} {\bibinfo {title}
  {{{MINFLUX}} nanometer-scale {{3D}} imaging and microsecond-range tracking on
  a common fluorescence microscope},\ }\href
  {https://doi.org/10.1038/s41467-021-21652-z} {\bibfield  {journal} {\bibinfo
  {journal} {Nature Communications}\ }\textbf {\bibinfo {volume} {12}},\
  \bibinfo {pages} {1478} (\bibinfo {year} {2021})}\BibitemShut {NoStop}%
\bibitem [{\citenamefont {Sahl}\ \emph {et~al.}(2024)\citenamefont {Sahl},
  \citenamefont {Matthias}, \citenamefont {Inamdar}, \citenamefont {Weber},
  \citenamefont {Khan}, \citenamefont {Br{\"u}ser}, \citenamefont {Jakobs},
  \citenamefont {Becker}, \citenamefont {Griesinger}, \citenamefont
  {Broichhagen},\ and\ \citenamefont {Hell}}]{sahl_direct_2024}%
  \BibitemOpen
  \bibfield  {author} {\bibinfo {author} {\bibfnamefont {S.~J.}\ \bibnamefont
  {Sahl}}, \bibinfo {author} {\bibfnamefont {J.}~\bibnamefont {Matthias}},
  \bibinfo {author} {\bibfnamefont {K.}~\bibnamefont {Inamdar}}, \bibinfo
  {author} {\bibfnamefont {M.}~\bibnamefont {Weber}}, \bibinfo {author}
  {\bibfnamefont {T.~A.}\ \bibnamefont {Khan}}, \bibinfo {author}
  {\bibfnamefont {C.}~\bibnamefont {Br{\"u}ser}}, \bibinfo {author}
  {\bibfnamefont {S.}~\bibnamefont {Jakobs}}, \bibinfo {author} {\bibfnamefont
  {S.}~\bibnamefont {Becker}}, \bibinfo {author} {\bibfnamefont
  {C.}~\bibnamefont {Griesinger}}, \bibinfo {author} {\bibfnamefont
  {J.}~\bibnamefont {Broichhagen}},\ and\ \bibinfo {author} {\bibfnamefont
  {S.~W.}\ \bibnamefont {Hell}},\ }\bibfield  {title} {\bibinfo {title} {Direct
  optical measurement of intramolecular distances with angstrom precision},\
  }\href {https://doi.org/10.1126/science.adj7368} {\bibfield  {journal}
  {\bibinfo  {journal} {Science}\ }\textbf {\bibinfo {volume} {386}},\ \bibinfo
  {pages} {180} (\bibinfo {year} {2024})}\BibitemShut {NoStop}%
\bibitem [{\citenamefont {Scheiderer}\ \emph {et~al.}(2025)\citenamefont
  {Scheiderer}, \citenamefont {Marin},\ and\ \citenamefont
  {Ries}}]{scheiderer_minflux_2025}%
  \BibitemOpen
  \bibfield  {author} {\bibinfo {author} {\bibfnamefont {L.}~\bibnamefont
  {Scheiderer}}, \bibinfo {author} {\bibfnamefont {Z.}~\bibnamefont {Marin}},\
  and\ \bibinfo {author} {\bibfnamefont {J.}~\bibnamefont {Ries}},\ }\bibfield
  {title} {\bibinfo {title} {{{MINFLUX}} achieves molecular resolution with
  minimal photons},\ }\href {https://doi.org/10.1038/s41566-025-01625-0}
  {\bibfield  {journal} {\bibinfo  {journal} {Nature Photonics}\ }\textbf
  {\bibinfo {volume} {19}},\ \bibinfo {pages} {238} (\bibinfo {year}
  {2025})}\BibitemShut {NoStop}%
\bibitem [{\citenamefont {Scheiderer}\ \emph {et~al.}(2024)\citenamefont
  {Scheiderer}, \citenamefont {{von der Emde}}, \citenamefont {Hesselink},
  \citenamefont {Weber},\ and\ \citenamefont {Hell}}]{scheiderer_minsted_2024}%
  \BibitemOpen
  \bibfield  {author} {\bibinfo {author} {\bibfnamefont {L.}~\bibnamefont
  {Scheiderer}}, \bibinfo {author} {\bibfnamefont {H.}~\bibnamefont {{von der
  Emde}}}, \bibinfo {author} {\bibfnamefont {M.}~\bibnamefont {Hesselink}},
  \bibinfo {author} {\bibfnamefont {M.}~\bibnamefont {Weber}},\ and\ \bibinfo
  {author} {\bibfnamefont {S.~W.}\ \bibnamefont {Hell}},\ }\bibfield  {title}
  {\bibinfo {title} {{{MINSTED}} tracking of single biomolecules},\ }\href
  {https://doi.org/10.1038/s41592-024-02209-6} {\bibfield  {journal} {\bibinfo
  {journal} {Nature Methods}\ }\textbf {\bibinfo {volume} {21}},\ \bibinfo
  {pages} {569} (\bibinfo {year} {2024})}\BibitemShut {NoStop}%
\bibitem [{\citenamefont {Weber}\ \emph {et~al.}(2021)\citenamefont {Weber},
  \citenamefont {Leutenegger}, \citenamefont {Stoldt}, \citenamefont {Jakobs},
  \citenamefont {Mihaila}, \citenamefont {Butkevich},\ and\ \citenamefont
  {Hell}}]{weber_minsted_2021}%
  \BibitemOpen
  \bibfield  {author} {\bibinfo {author} {\bibfnamefont {M.}~\bibnamefont
  {Weber}}, \bibinfo {author} {\bibfnamefont {M.}~\bibnamefont {Leutenegger}},
  \bibinfo {author} {\bibfnamefont {S.}~\bibnamefont {Stoldt}}, \bibinfo
  {author} {\bibfnamefont {S.}~\bibnamefont {Jakobs}}, \bibinfo {author}
  {\bibfnamefont {T.~S.}\ \bibnamefont {Mihaila}}, \bibinfo {author}
  {\bibfnamefont {A.~N.}\ \bibnamefont {Butkevich}},\ and\ \bibinfo {author}
  {\bibfnamefont {S.~W.}\ \bibnamefont {Hell}},\ }\bibfield  {title} {\bibinfo
  {title} {{{MINSTED}} fluorescence localization and nanoscopy},\ }\href
  {https://doi.org/10.1038/s41566-021-00774-2} {\bibfield  {journal} {\bibinfo
  {journal} {Nature Photonics}\ }\textbf {\bibinfo {volume} {15}},\ \bibinfo
  {pages} {361} (\bibinfo {year} {2021})}\BibitemShut {NoStop}%
\bibitem [{\citenamefont {Wirth}\ \emph {et~al.}(2023)\citenamefont {Wirth},
  \citenamefont {Scheiderer}, \citenamefont {Engelhardt}, \citenamefont
  {Engelhardt}, \citenamefont {Matthias},\ and\ \citenamefont
  {Hell}}]{wirth_minflux_2023}%
  \BibitemOpen
  \bibfield  {author} {\bibinfo {author} {\bibfnamefont {J.~O.}\ \bibnamefont
  {Wirth}}, \bibinfo {author} {\bibfnamefont {L.}~\bibnamefont {Scheiderer}},
  \bibinfo {author} {\bibfnamefont {T.}~\bibnamefont {Engelhardt}}, \bibinfo
  {author} {\bibfnamefont {J.}~\bibnamefont {Engelhardt}}, \bibinfo {author}
  {\bibfnamefont {J.}~\bibnamefont {Matthias}},\ and\ \bibinfo {author}
  {\bibfnamefont {S.~W.}\ \bibnamefont {Hell}},\ }\bibfield  {title} {\bibinfo
  {title} {{{MINFLUX}} dissects the unimpeded walking of kinesin-1},\ }\href
  {https://doi.org/10.1126/science.ade2650} {\bibfield  {journal} {\bibinfo
  {journal} {Science}\ }\textbf {\bibinfo {volume} {379}},\ \bibinfo {pages}
  {1004} (\bibinfo {year} {2023})}\BibitemShut {NoStop}%
\bibitem [{\citenamefont {Deguchi}\ and\ \citenamefont
  {Ries}(2024)}]{deguchi_simple_2024}%
  \BibitemOpen
  \bibfield  {author} {\bibinfo {author} {\bibfnamefont {T.}~\bibnamefont
  {Deguchi}}\ and\ \bibinfo {author} {\bibfnamefont {J.}~\bibnamefont {Ries}},\
  }\bibfield  {title} {\bibinfo {title} {Simple and robust {{3D MINFLUX}}
  excitation with a variable phase plate},\ }\href
  {https://doi.org/10.1038/s41377-024-01487-1} {\bibfield  {journal} {\bibinfo
  {journal} {Light: Science \& Applications}\ }\textbf {\bibinfo {volume}
  {13}},\ \bibinfo {pages} {134} (\bibinfo {year} {2024})}\BibitemShut
  {NoStop}%
\bibitem [{\citenamefont {Bezanson}\ \emph {et~al.}(2017)\citenamefont
  {Bezanson}, \citenamefont {Edelman}, \citenamefont {Karpinski},\ and\
  \citenamefont {Shah}}]{bezanson_julia_2017}%
  \BibitemOpen
  \bibfield  {author} {\bibinfo {author} {\bibfnamefont {J.}~\bibnamefont
  {Bezanson}}, \bibinfo {author} {\bibfnamefont {A.}~\bibnamefont {Edelman}},
  \bibinfo {author} {\bibfnamefont {S.}~\bibnamefont {Karpinski}},\ and\
  \bibinfo {author} {\bibfnamefont {V.~B.}\ \bibnamefont {Shah}},\ }\bibfield
  {title} {\bibinfo {title} {Julia: {{A Fresh Approach}} to {{Numerical
  Computing}}},\ }\href {https://doi.org/10.1137/141000671} {\bibfield
  {journal} {\bibinfo  {journal} {SIAM Review}\ }\textbf {\bibinfo {volume}
  {59}},\ \bibinfo {pages} {65} (\bibinfo {year} {2017})}\BibitemShut {NoStop}%
\bibitem [{\citenamefont {Mogensen}\ and\ \citenamefont
  {Riseth}(2018)}]{mogensen_optim_2018}%
  \BibitemOpen
  \bibfield  {author} {\bibinfo {author} {\bibfnamefont {P.~K.}\ \bibnamefont
  {Mogensen}}\ and\ \bibinfo {author} {\bibfnamefont {A.~N.}\ \bibnamefont
  {Riseth}},\ }\bibfield  {title} {\bibinfo {title} {Optim: {{A}} mathematical
  optimization package for {{Julia}}},\ }\href
  {https://doi.org/10.21105/joss.00615} {\bibfield  {journal} {\bibinfo
  {journal} {Journal of Open Source Software}\ }\textbf {\bibinfo {volume}
  {3}},\ \bibinfo {pages} {615} (\bibinfo {year} {2018})}\BibitemShut {NoStop}%
\end{thebibliography}%

\vspace*{2ex}
\appendix
\section{Implementation details}
A proof-of-principle implementation of the Bayesian approach presented in the Julia programming language \cite{bezanson_julia_2017} here is available at \href{https://gitlab.gwdg.de/sschult/minflux}{\url{https://gitlab.gwdg.de/sschult/minflux}}.

The posterior $P_k$ was approximated on a uniform Cartesian grid with positions $\vec x_{ij} = (x_i, y_j)$ and associated probabilities $p_{ij}$.
The Bayesian updates were performed according to eq.~\eqref{eq: bayes} and after each step the posterior was normalized such that $\sum_{ij} p_{ij} = 1$. 
The initial grid consisted of $60\times60$ points with $15\,\mathrm{nm}$ spacing. 
After each update, the grid was adaptively subdivided to maintain a grid spacing of at most $\operatorname{std}(P_k)/10$.
Probabilities at new grid points were obtained by linear interpolation.
To improve efficiency, grid positions with negligible probability mass were pruned: all $x_i$ with $\max_j p_{ij} < 10^{-9}$ and all $y_j$ with $\max_i p_{ij} < 10^{-9}$ were removed.

For evaluation of the expected information gain (eq.~\eqref{eq: eig}) the infinite sum over the photon counts $n$ was truncated at $n = \mathrm{max}(5, \lceil\mu + 50\sqrt{\mu}\rceil)$.
The optimal donut placement (eq.~\eqref{eq: argmaxeig}) was determined using the L-BFGS optimization algorithm as implemented in the Optim.jl package \cite{mogensen_optim_2018}.

\end{document}